\documentclass[reprint,onecolumn,superscriptaddress,a4paper,nofootinbib]{revtex4-1}
\usepackage{graphicx}
\usepackage{footnote}
\usepackage{bm}
\usepackage{amsmath}
\usepackage{amsfonts}
\usepackage{microtype}
\usepackage{textcomp}
\usepackage{mathtools}
\usepackage{lineno}
\usepackage{chemfig}
\usepackage{palatino}
\usepackage{soul}
\usepackage{cancel}
\usepackage[]{subfig}
\DeclareCaptionLabelSeparator{frenchsep}{ : }
\usepackage{mwe}
\usepackage[section]{placeins}
\usepackage[font={small}]{caption}

\begin{document}

\title{A cryogenic magneto-optical device for long wavelength radiation}
\author{S.J. Rezvani$^*$}
\affiliation{Laboratori Nazionali di Frascati, INFN, Via Enrico Fermi, 00054 Frascati, Italy.}
\affiliation{Consiglio Nazionale delle Ricerche (CNR), CNR-IOM, 34149 Basovizza, Italy.}
\email{corresponding author. Email:rezvani@lnf.infn.it}
\author{D. Di Gioacchino}
\affiliation{Laboratori Nazionali di Frascati, INFN, Via Enrico Fermi, 00054 Frascati, Italy.}
\author{S. Tofani}
\affiliation{Department of Information Engineering, Electronics and Telecommunications, "Sapienza" University of Rome, 00184 Rome, Italy.}
\author{A. D'Arco}
\affiliation{Department of Physics, Sapienza University of Rome, 00185 Rome, Italy.}
\affiliation{INFN - Roma1, P.le Aldo Moro 2, 00185 Rome, Italy.}
\author{C. Ligi}
\affiliation{Laboratori Nazionali di Frascati, INFN, Via Enrico Fermi, 00054 Frascati, Italy.}
\author{S. Lupi}
\affiliation{Department of Physics, Sapienza University of Rome, 00185 Rome, Italy.}
\affiliation{Laboratori Nazionali di Frascati, INFN, Via Enrico Fermi, 00054 Frascati, Italy.}
\author{C. Gatti}
\affiliation{Laboratori Nazionali di Frascati, INFN, Via Enrico Fermi, 00054 Frascati, Italy.}
\author{M. Cestelli Guidi}
\affiliation{Laboratori Nazionali di Frascati, INFN, Via Enrico Fermi, 00054 Frascati, Italy.}
\author{A. Marcelli}
\affiliation{Laboratori Nazionali di Frascati, INFN, Via Enrico Fermi, 00054 Frascati, Italy.}
\begin{abstract}
We present here a small-scale liquid Helium (LHe) immersion cryostat with an innovative optical setup suitable to work in long wavelength radiation ranges and under applied magnetic field. The cryostat is a multi stage device with several shielding in addition to several optical stages. The system has been designed with an external liquid Nitrogen boiler to reduce the liquid bubbling. The optical and mechanical properties of the optical elements were calculated and optimized for the designed configuration while the optical layout has been simulated and optimized among different configurations based on the geometry of the device. The final design has been optimized for low noise radiation measurements of proximity junction arrays under applied magnetic field in the wavelength range $\lambda$=250-2500 $\mu$m.
\end{abstract}
\maketitle

\section{Introduction}
Cryogenic temperatures are required in many researches such as superconductivity and superfluidity, surface and interface, advanced spectroscopies and in the R\&D of low-noise detectors. As an example, superconducting radiation detectors integrated into surface Paul traps for scalable quantum information processing require operation at cryogenic temperatures \cite{Niedermayr_2014,schwarz70,micke93}. Optical spectroscopies at low temperature usually allow to resolve narrow features and trap unstable intermediates \cite{andrews43}.
Recently, long wavelength radiations have come to attention being a particular region of the electromagnetic spectrum interesting for multi disciplinary applications for both basic researches and technologies \cite{RN1523,Ortolani2006,Nanni2015}. This e.m. domain $\lambda$=250-2500 $\mu$m (energy range 0.5-4 meV) allows the investigation of several fundamental physical phenomena, e.g., phonon and plasmon dynamics, elementary particle physics and possibly also cold dark matter \cite{Jepsen2011,Saeedkia2013,Karasik2011,Toma2015,DApuzzo2017}. Furthermore, with its high transmission through a wide range of non-conducting materials, long wavelength radiation and, in particular terahertz (THz) radiation, holds a significant potential in several applications.
%However, in comparison with adjacent energy ranges, e.g., IR and UV, in this wavelength domain the instrumentation and, in particular, detectors are still limited, expensive or difficult to use. Moreover, instrumentation and detectors are limited in this wavelength region, expensive and/or difficult to use compared to similar devices operating in the Infrared (IR) and ultraviolet (UV) energy regions. For instance semiconducting bolometers are widely used as cryogenic and non-cryogenic terahertz detectors \cite{richards2}, but their sensitivity to temperature and mechanical fluctuations and their limitations to the high frequency band operation, renders their use mainly confined to dedicated cryogenic laboratories. Recent studies have shown that also superconducting devices can be successfully employed at long wavelengths and in particular for the generation and detection of far-IR and THz radiation \cite{Ozyuzer2007,Du2015,Hammar2011,Seliverstov2015}. These devices exhibit an extremely low noise compared to the semiconducting counterparts, a response time orders of magnitude lower and a higher frequency range of operation.

In the recent years, the investigation of semiconducting and superconducting behavior of many materials and the modulation of the electronic properties of systems with low dimensionality have been investigated \cite{Rezvani_2016_3,Rezvani2,Rezvani_2016,Rezi_4}.
It has been demonstrated that superconducting proximity junction arrays can be designed to behave as long wavelength radiation detectors \cite{Rezvani2020,Gioacchino2017}. The superconducting dynamics of such devices, based on non-localized vortex dynamics, can be modulated via external magnetic and electric fields \cite{Poccia2015}. The response of these devices in the radio frequency domain, well below the expected plasma frequency of the proximity junction, can be tuned via non-equilibrium phenomena induced by external perturbations. It has been also speculated that such devices may represent a new pathway towards robust, low noise, broadband and with high sensitivity low energy detectors. However, the investigation of the dynamics in such devices requires a low noise cryostat with a highly stable temperature control that can be operated under external magnetic and electric fields.
These cryogenic systems should also be designed with suitable optical arrangements compatible with this energy range. In a conventional cold finger optical system, the sample is cooled through its mechanical link to the cooling element, either by boiling liquid Helium or a mechanical refrigerator. This approach presents several problems. The first is that the sample is cooled relatively slowly and often we observe the formation of crystallites on the sample surface. In addition, reaching the required temperatures and its fine control under applied electric and magnetic fields is not an easy task. Finally, the optimization of the optical arrangement to operate in this extreme photon energy range is complex.

Within the framework of the project TERA \cite{Tera}, we have developed a small scale liquid Helium immersion cryostat with an embedded optical setup, aimed at the detection of the photo-response of novel proximity junction arrays under an applied magnetic field. In the following we will present and discuss the different cooling stages, the shields and the optical layout. The latter has been simulated and optimized based on the different in-operando configurations.
The optical transmission and the mechanical resistance of the optical components have been calculated, optimized and characterized.
\section{Cryostat and optical set up description}
\begin{figure}[h!]
\centering
\includegraphics[width=0.5\textwidth]{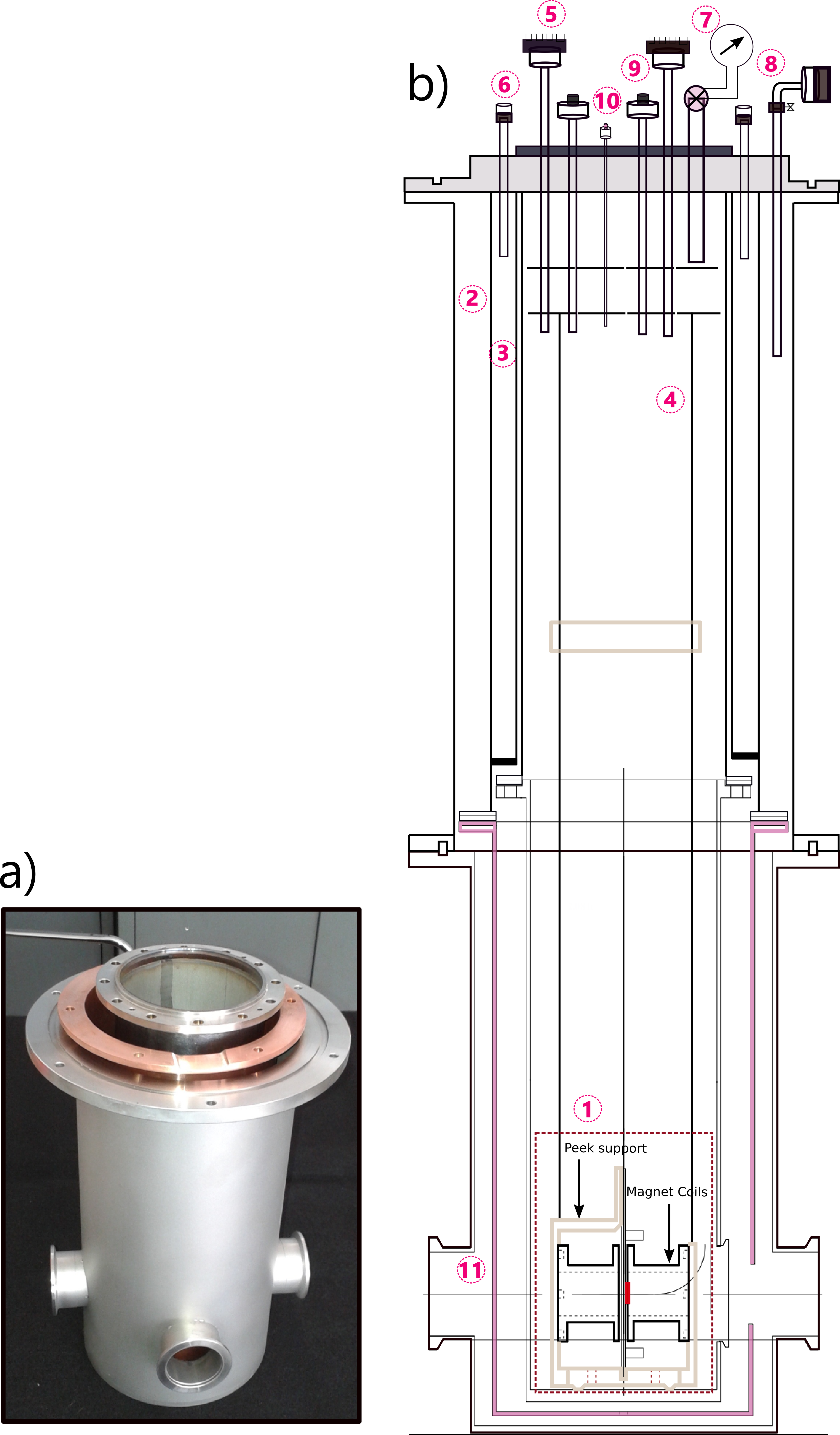}
\caption{a) image of the lower half of the cryostat showing the vacuum jacket, the copper shield and the inner LHe container; b) Section of the optical cryostat. 1) sample compartment inside the opto-magnetic set up; 2) the vacuum jacket; 3) the liquid Nitrogen reservoir; 4) the liquid Helium reservoir 5) connector for signal, thermometer and gauss-meter wiring; 6) liquid Nitrogen insert tube ; 7) liquid Helium pressurizing and pumping valve; 8) vacuum jacket pumping valve; 9) liquid Helium insert tube; 10) LHe level sensor feedthrough; 11) copper radiation shield. Inside the inner tube there are two copper disks as radiation shields. The sample holder is connected to the lower one via two Rexilon rods to minimize thermal losses.}
\label{cryo}
\end{figure}

A schematic view of the immersion cryostat is shown in Fig.\ref{cryo}. The cryostat is designed to be a table top small scale cryostat for facile optical alignments. The section views of the cryostat (Fig. \ref{cryo}b) shows the vacuum chamber (2) surrounding the liquid Nitrogen tank (3) in the upper half, and the polished copper radiation shield in the lower one (11). The shield has a hole in correspondence of the optical path. At the center, the liquid Helium container (4), inside which the experimental apparatus (1) is hosted. The inner container has a 10 cm diameter, a value that sets the maximum size of both magnetic coils and sample holder. The inner stainless steel tube is provided at the bottom with a window assembly mounted on a CF flange that holds a 29 mm diameter optical window set in front of the sample.

The copper cooling jacket extends down the sides as well as below the window assembly to improve cooling and to speed up the initial cooling process. The external container is equipped with four vacuum-tight optical windows. All inner metal-metal connections, except the copper shield and the cryostat cover part that is connected by bolts, have been connected either by soldering or brazing. Inside upper part of the liquid Helim container are housed two concentric copper disks acting as thermal shields and connected to the sample holder support with two Rexilon rods (10 mm diameter), useful to minimize the thermal conduction between the immersed components and the thermal shields.

\begin{figure}[h!]
    \includegraphics[width=0.5\textwidth]{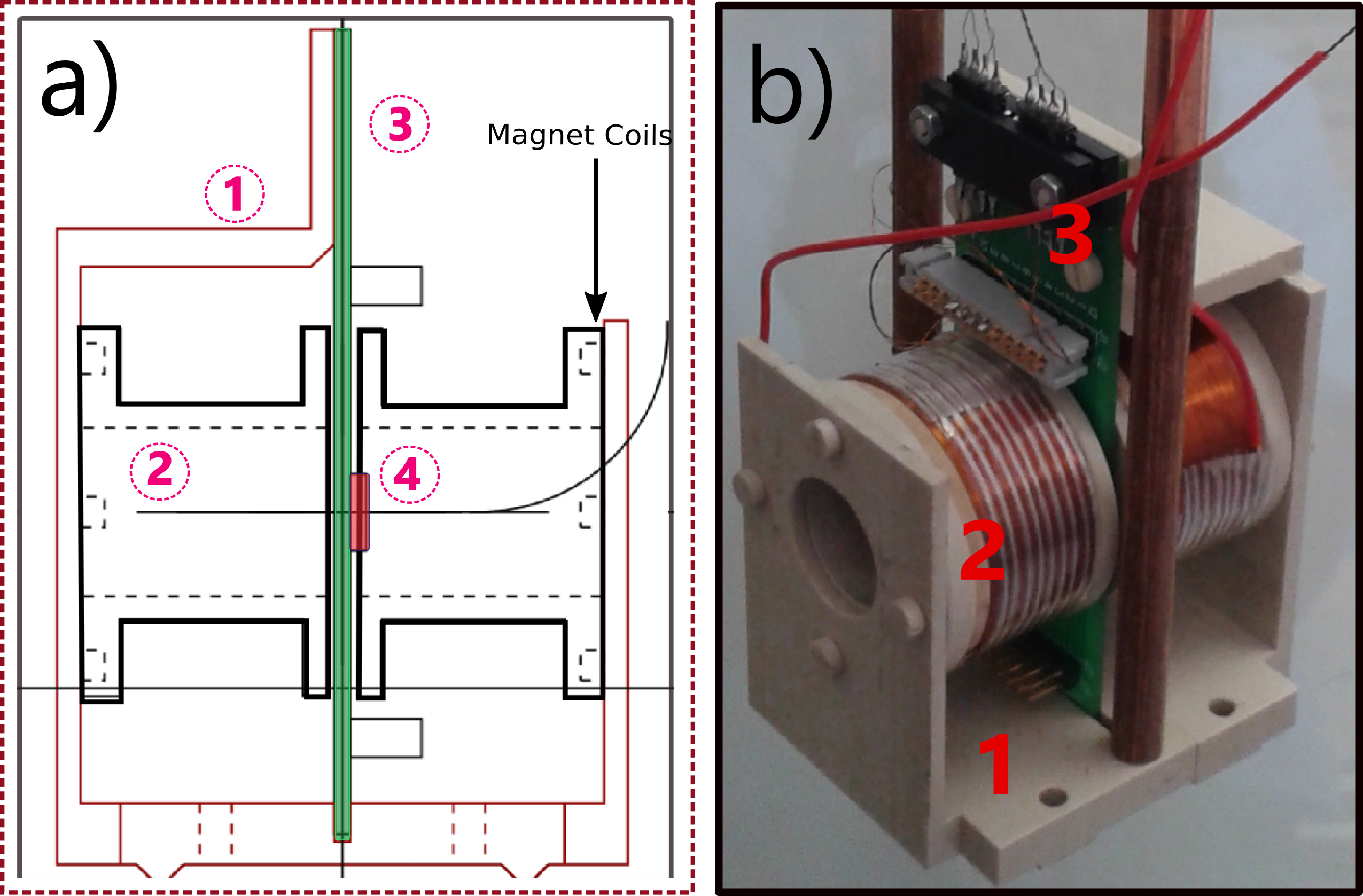}
    \caption{a)  The layout of the sample holder: (1) the PEEK sample holder; (2) the superconducting magnetic coils made by NbN wires; (3) PCB sample support and electrical contacts and (4) the sample placed in the center. b) image of the sample holder.}
    \label{holder}
    \end{figure}

The sample holder support consists of a PEEK holder support housing two superconducting NbN coils attached with concentric tubes with 21 mm apertures pointing to the sample set in the center. Considering the aperture dimension of the magnetic coils the maximum sample size can be 20$\times$ 20 mm and can be replace by simply opening the cryosts top flange (top dark part of figure\ref{cryo}b), since the  sample holder and the magnetic coils are connected to the copper disks shields and connected to this flange.  The coils generate a magnetic field normal to the sample surface plane (see Fig. \ref{holder}) up to 150 mT at 4.2 K calibrated with a calibrated gaussmeter.
Under operation the vacuum jacket chamber reaches up to $1 \times 10^{-4}$ mbar rapidly while the vacuum level lowers to $5 \times 10^{-6}$ mbar, introducing the liquid Nitrogen inside the LN bath. The latter should be refilled periodically prior to introduction of the liquid Helium. The thermal conduction from the bottom and the contact thermal conduction lead to a temperature of $\sim 100$ K in the main cylinder upon insertion of the liquid Nitrogen. The LHe insertion takes place in over pressurized He inner cryotube to prevent the crystallization on the sample surface. The temperature is read by a calibrated Cernox thermometer connected to the PCB sample support and the applied magnetic field is measured via a previously calibrated Hall probe inserted perpendicularly inside the coil at the back of sample. The cryostat with the capacity of the $\sim 5$ Lt, is stable and maintains the temperature to 4.2 K for about six hours without incident radiations. The LN evaporator jacket reduces bubbling inside the cryotube a condition that significantly reduces the light scattering along the optical path \cite{andrews43}. The temperature can be further reduced down to $\sim $2 K with a continuous pumping of the pressurized LHe bath.

\begin{figure}[h!]
\centering
\includegraphics[width=\textwidth]{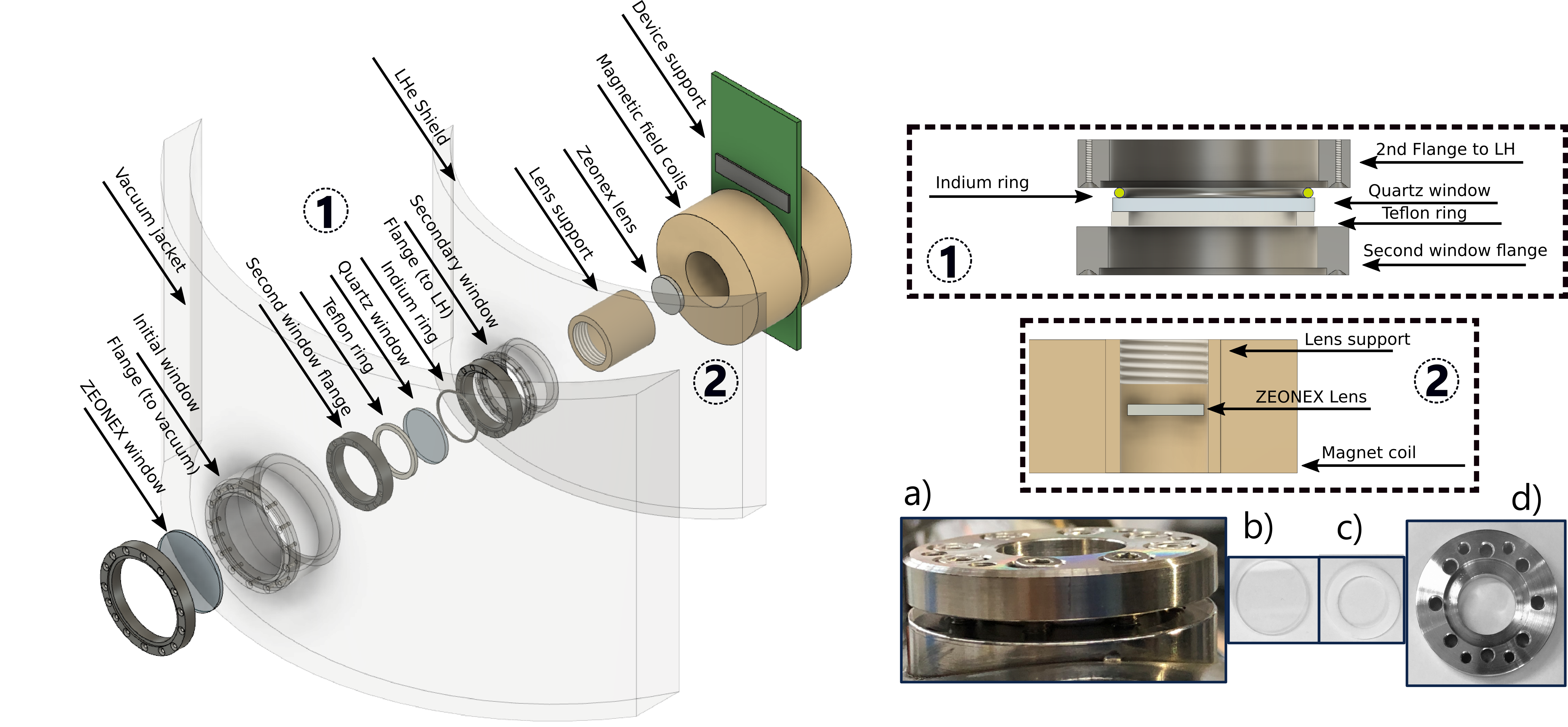}
\caption{Layout of the optical set up from the outer vacuum jacket of the cryostat to the inner coil lens positioning. The two segments indicated by number 1 and 2 are shown disassembled and in the insets after the assembly. a) The actual configuration as indicated in frame (1); b) the optical lens used in the layout ; c) the Teflon ring; d) the modified Conflat Flange.}
\label{optic}
\end{figure}

The detailed cryogenic optical arrangement is described in figure \ref{optic}. The layout consists of a 2 mm Zeonex window mounted with a Vyton O-ring to the vacuum chamber jacket and connected to the ambient temperature. Along the optical path the second window is in contact with the LHe under a heavy tensile pressure (due to the differential pressure between the two volumes). The window is made by 2.2 mm thick crystal quartz as will be discussed later. The window is connected to the LHe cryotube by a modified Conflat flange and the vacuum is sealed by a teflon ring on the vacuum side and an Indium wire on the liquid side (see figure \ref{optic}-inset 1). The window was tightened in gradual cross contraction using a torque driver up to 2 N/m.

A 2 mm diameter Zeonex focusing lens with a 20 mm focal distance is set in from of the sample, screwed within the coil aperture inside the PEEK cylinder support (see figure \ref{optic}-inset 2). The cylinder itself is fixed by a PEEK nut tightening the cylinder in place. This design further facilitates the setting along the axis of the lens focal point with regard to the sample positioning.

\begin{figure}[h!]
    \centering
    \includegraphics[width=0.4\textwidth]{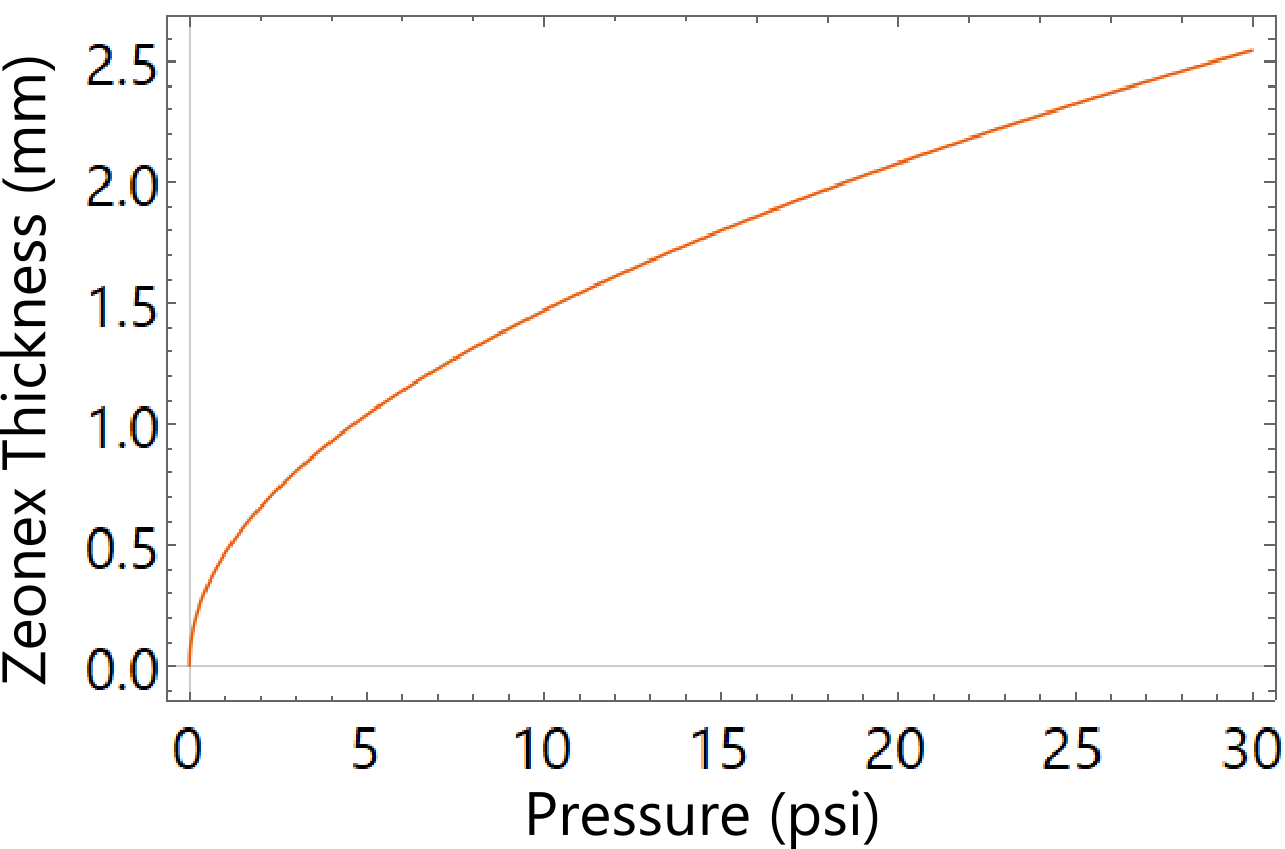}
    \caption{The curve of the Zeonex window slab thickness as a function of the pressure difference between the cryostat jackets.}
    \label{thickness}
    \end{figure}

The optimal thickness of all windows, being the minimum thickness capable of withstanding the pressure difference insuring its high transmittance, was calculated considering the materials tensile strengths as
\begin{equation}
    d=\sqrt{\frac{1.1\times P \times R^2 \times S}{\eta}}
\end{equation}
where, $d$ is the slab thickness, $R$ is the unsupported area, $P$ is the pressure difference, $S$ the safety factor and $\eta$ is the modulus of the rupture. The relative thickness behavior vs. differential pressure of the crystal quartz with $\eta = 9427$ psi with a maximum 3 mm unsupported area is shown in figure \ref{thickness}. Considering the differential pressure to sustain, a minimum thickness in the range $d= 1.8-2.2$ mm was considered.
While for the outer Zeonex window with $\eta=6090$ psi  withstanding the atmospheric pressure a relatively smaller thickness of the windows has been considered (2 mm). Both Zeonex and crystal quartz windows were tested under these conditions and passed the tests. The Zeonex window was tested also under vacuum down to $1 \times 10^{-6}$ mbar.

\begin{figure}[h!]
    \centering
    \includegraphics[width=0.9\textwidth]{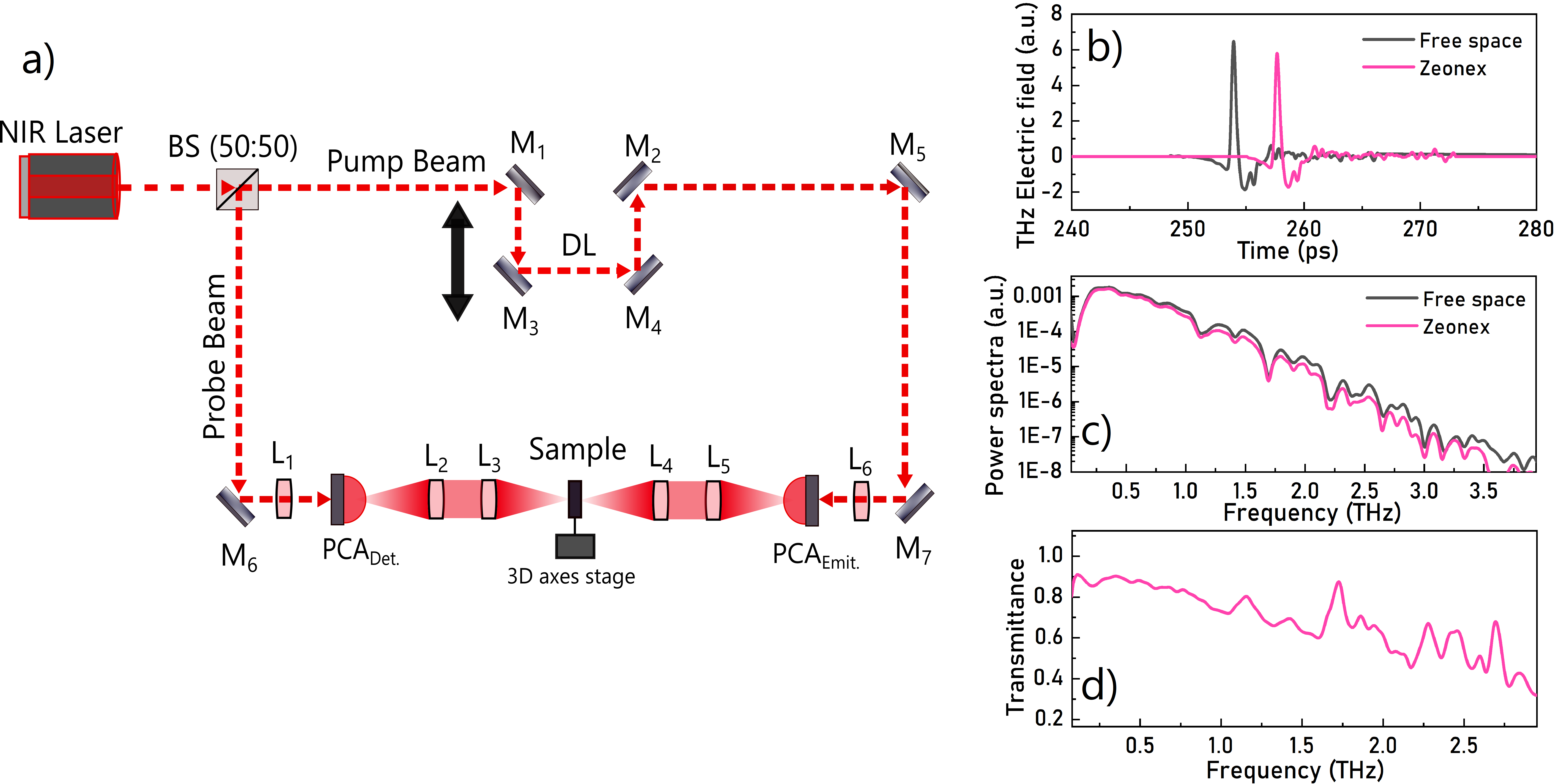}
    \caption{a) THz-TDS system in transmission geometry for spectroscopic characterization of the Zeonex window. b) THz electric fields: THz electric field relative to free space (black line) and THz transmitted electric field of the Zeonex window (red line); c) Power spectra relative to free space (black line) and 2 mm thick Zeonex window (red line); d) Transmittance of 2 mm thick Zeonex window slab.}
    \label{THZ}
    \end{figure}

%The optical response of the Zeonex window was measured in the required frequency range to evaluate the effective transmission of the windows set and consequently the efficiency of the optical layout.
In order to evaluate the effective transmission and the efficiency of the optical layout, the optical response of the Zeonex windows were measured in the spectral region of interest.
The optical response of the polymeric Zeonex material, was investigated with terahertz time-domain spectroscopy (THz-TDS), described in detail in Refs\cite{Castro-Camus16,Ferguson2002,Zhang2010}. The in house THz-TDS spectrometer in the transmission configuration, used for this purpose, is based on two photo-conductive switches, one as emitter of the broadband THZ radiation (0.1-2.5 THz) and one as receiver \cite{anna}, respectively (see figure \ref{THZ}a). Two twin G10620-11 Hamamatsu photo-conductive antennas (PCAs) were operated by a mode-locked ultrafast laser (FemtoFiber NIRpro, Toptica), at 780 nm, temporal pulse width 100 fs, repetition rate of 80 MHz and output power of 150 mW. After the laser power reduction, a beam splitter (BS) 50:50 splits the main beam into two separated pump and probe beams. Dielectric mirrors propagate the beams towards emitter and receiver PCAs. The THz produced by the emitter, is then collected and collimated by TPX lenses with focal length of 50 mm. The THz beam is focused on the sample and the transmitted radiation is recollected, re-collimated and refocused on the receiver PCA. Simultaneously, the probe beam is used to gate the THz-detector PCA. Two motorized stages are used in the system configuration (see figure \ref{THZ}). The fist one, a delay line (DDSM100/M, Thorlabs) allows the collection of temporal THz electric field varying the optical path between pump and probe beams \cite{Castro-Camus16}.
The second one is a 3-axes motor stage that ensures an accurate positioning and alignment of the sample in the THz beam focal spot.

The output signal from the THz-detector, filtered by a lock-in amplifier, was acquired and digitized and then transferred to the PC. The optical system provides a spectral bandwidth ranging from 0.1 to 2.5 THz, with a spectral resolution of 50 GHz. This spectral resolution is sufficient to characterize the response of the non-resonant and transparent window set.  We recorded 5 THz electric field profiles with 5 scans. The THz electric fields for the free space (without sample) and for the Zeonex window with thickness 2 mm are shown in figure \ref{THZ}b. The temporal delay between the air and the Zeonex signal was $3.75\pm 0.01$ ps. The power spectra of free space and Zeonex slab are also shown in figure \ref {THZ}c along with its total transmittance (see Fig. \ref{THZ}d).
The transmittances was calculated as
\begin{equation}
T_{Zeonex} = \frac{\left|E_{Zeonex}(\omega)\right|^2}{\left|E_{Freespace}(\omega)\right|^2}
\end{equation}
The results indicate a transmission of about 80\% in the range below 1 THz, which is slightly higher than that of crystal quartz \cite{Cha2006}.

\begin{figure}[h!]
    \centering
    \includegraphics[width=0.4\textwidth]{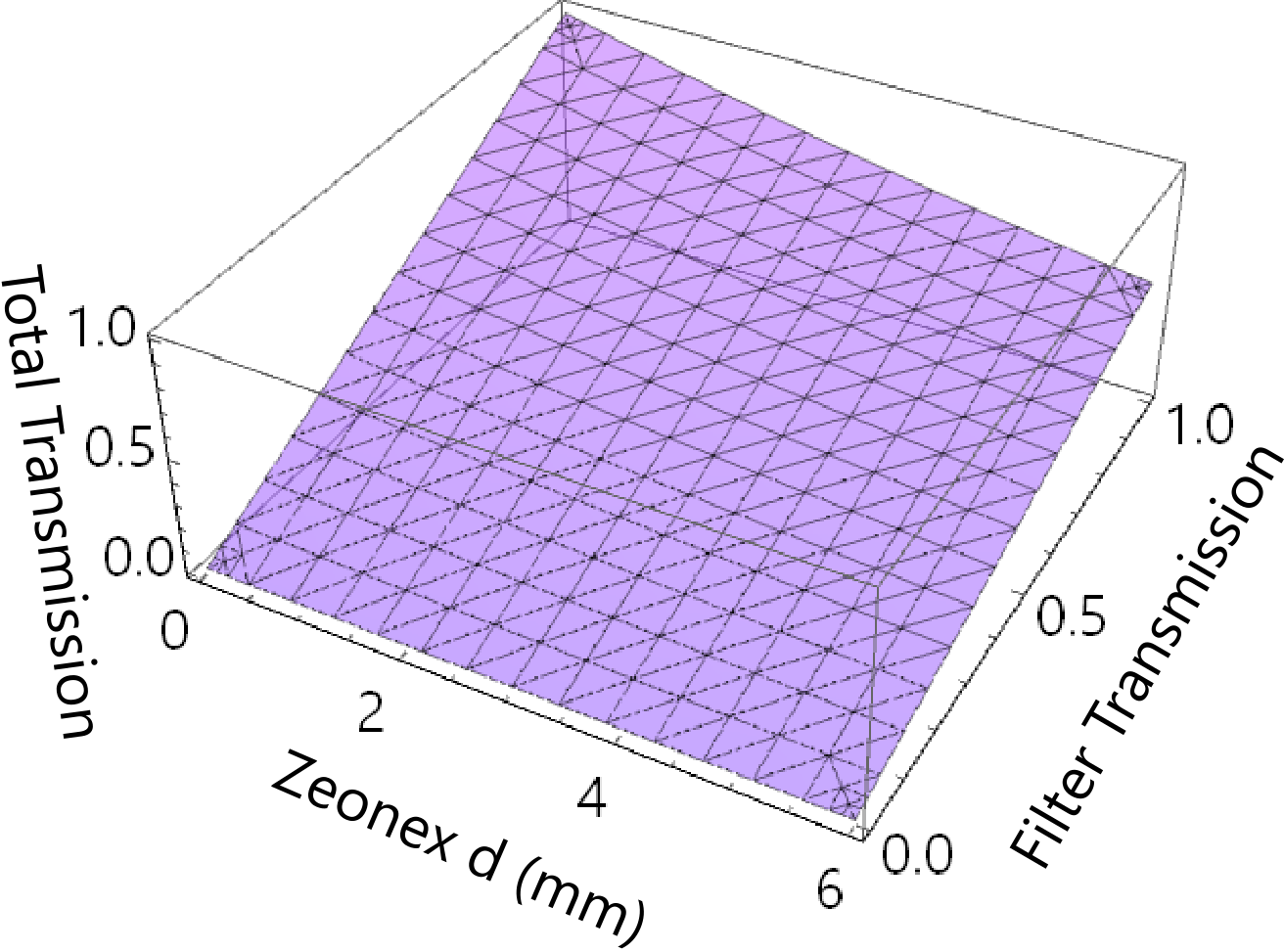}
    \caption{The total transmission of the optical layout with the Zeonex windows and the IR-Vis filter.}
    \label{trans}
    \end{figure}

The total transmission of the optical arrangement was then calculated using the Beer-Lambert law \cite{Born2000} and considering the transmission of the Zeonex derived from experimental results (see figure \ref{trans}). In order to prevent any contamination of visible and IR radiation inside the cryotube a 0.2 mm flexible HDPE filter (with T$\sim$ 85\% at 1 THz \cite{tydex}) was attached to the entrance window. Based on the thickness of the Zeonex windows, crystal quartz and the filter transmission a total transmission of about 45-50\% onto the sample was calculated.

\begin{figure}[h!]
    \centering
    \includegraphics[width=0.8\textwidth]{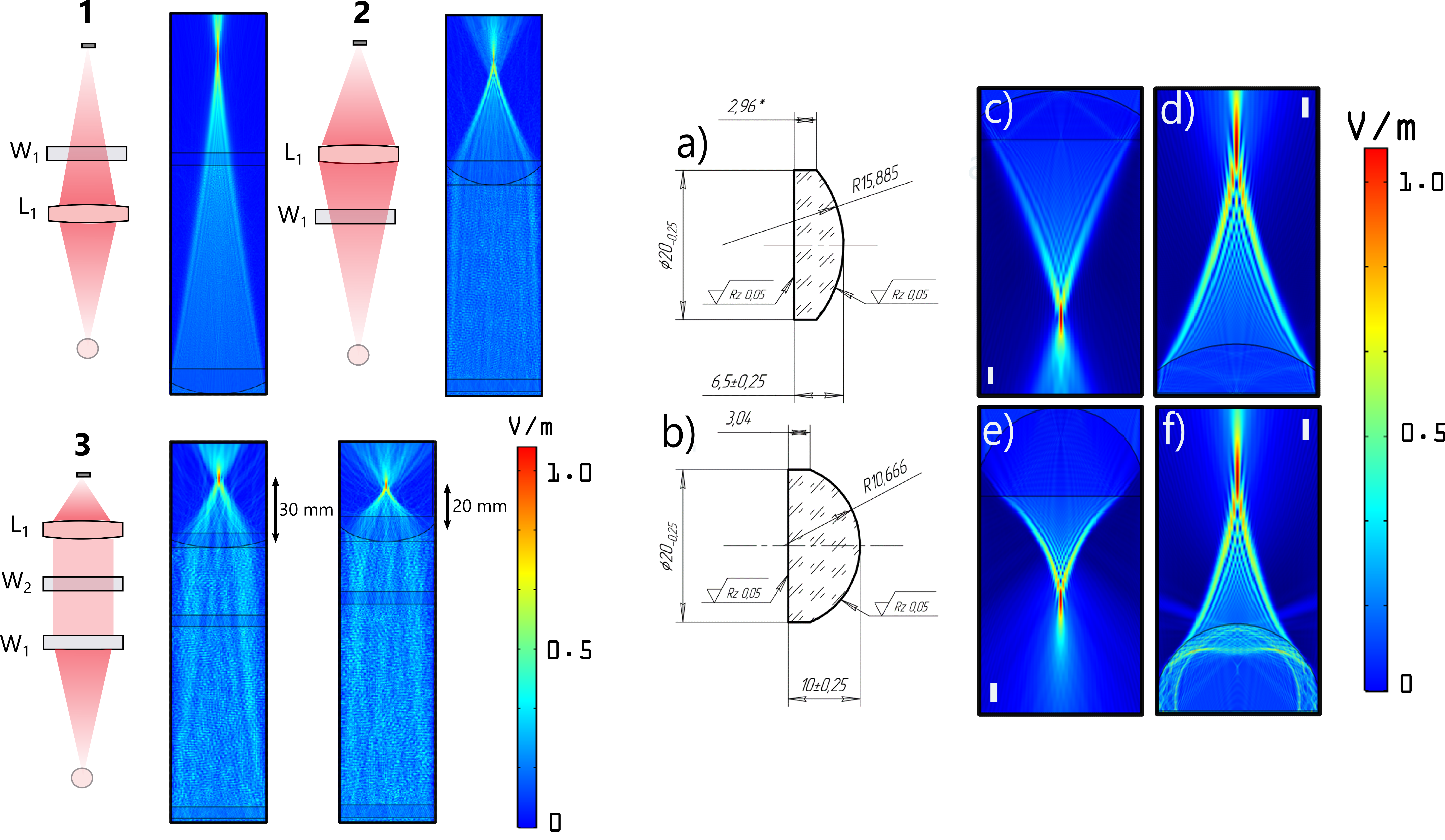}
    \caption{1-3) Different optical configurations based on the geometry of the the cryostat and their simulations; (1) a single outer jacket lens and a single inner window; (2) a single outer jacket window and a single inner tube lens; (3) an outer jacket window, an inner tube window and one very close lens ; a-f) different lenses configurations and their DoF calculation. The bars show the normalized modulus of the electric field.}
    \label{lens}
    \end{figure}

Based on the geometry of the cryogenic system, three different layouts of the optical paths were considered as described in Fig.\ref{lens}. All of them are possible configurations of this optical system. Key parameters, considered mainly in the lens design, were the resolution, the focal length and the lens diameter. Both focal length and lens diameter are determined by geometrical constraints.
%the focal length and resolution and the lens diameters, both bounded to the geometrical constraints.
Furthermore, we planned to have a focal spot diameter less than 1 mm in order to guarantee a sufficient number of photons on the sample. In the layout (1), a THz plane wave illuminates the focusing lens (L1) with a focal length of 130 mm. A polymeric/quartz window (W1) is located between the lens (L1) and its focal plane. The maximum diameter for the lens in this location is 50 mm. In the layout (2) the THz radiation pass the polymeric window (W1) and it is focused by the lens (L1) with a focal length of about 50 mm that acts also as the second window. In the layout (3), the plane wave is incident on two polymeric/quartz windows, (W1) and (W2), and it is finally focused by the lens (L1) with a focal length of about 20 mm (length of the magnetic coil aperture L=32 mm). Considering the constraints of the coil, the maximum lens diameter is 20 mm since the diameter of the magnetic coil aperture is D=21 mm.

All layouts have been numerically simulated by means of a commercial software based on the finite element method in the frequency domain \cite{comsol}. The material selected for both windows and lenses is the Zeonex by Zeon Corporation, a cyclo-olefin with a refractive index of $\epsilon=1.518-j0.001$ at 1 THz, successfully employed in other devices operating at THz frequencies \cite{Tofani2019,Tofani20192}. The windows have been designed with a thickness of 2 mm. The lens (L1) in the layout 1 is a plano-convex lens with a radius of curvature R = 67.6 mm and a maximum thickness of 1 cm. The lens (L1) in the layout (3) is a plano-convex lens with a radius of curvature R = 10.7 mm and a maximum thickness of 1 cm. The lens (L1) in the layout (2) is a plano-convex lens with a radius of curvature R = 26.2 mm and a maximum thickness of 1 cm. As a proof of principle, a linearly polarized plane wave has been selected as a radiation source at 1 THz.

The results of the numerical simulations are described in Fig.\ref{lens}. They confirm the occurrence of a focal spot with a diameter comparable with the theory \cite{Born2000}. The full width half maximum (FWHM) of the absolute value of the electric field in the focal area corresponds to a measure of the focal spot diameter of the selected lenses. The FWHM of the lens (L$_1$) focal spot is equal to about 1 mm for the layout (1), 0.48 mm for the layout (3), and 0.65 mm for the layout (2). As expected, the presence of the windows slightly affects the focusing properties in terms of maximum values of the electric field modulus at the focal point, due to the mismatch of the refractive index between vacuum and Zeonor and to the moderate absorption of the polymer at 1 THz.
The optical configuration presented in layout (3) shows a higher resolution of the lens (L$_1$) at 1 THz, compared with the other configurations. Hence, we selected the latter optical arrangement for the cryogenic system that was designed and manufactured \cite{tydex} according to this layout (3).

Furthermore, to implement the tuning of the lens depth of field (DoF), defined as the distance along the optical axis in which the power does not decrease below the 80\% of the maximum, a second lens with a focal length of 30 mm alternative to the lens (L1) of the Layout (3) was numerically simulated. The modulation of the DoF facilitates the compensation of the photon intensity attenuation due to sample positioning misalignments. The plano-convex lens with a focal length of 30 mm has a diameter of 20 mm, a radius of curvature R = 15.9 mm, and a maximum thickness of 6.5 mm. Simulations of this lens in layout (3) are reported in Fig.\ref{lens}. The FWHM of the resulting focal spot is equal to about 0.56 mm. Simulations confirm that a longer focal length allows a longer DoF.
%: the lens in the layout (1) (focal length equal to 130 mm) has a DoF of about 8 mm, while the lens in the layout (3) (focal length of 20 mm) has a DoF of about 2.3 mm. For this reason,
We also evaluated the lens orientation in term of DoF by numerical simulations. As reported in \ref{lens}, when the lens is oriented as in Fig.\ref{lens} c and e, DoF is 2.3 mm and 3.5 mm for the lens with a focal length of 20 mm and 30 mm, respectively. On the other hand, when the lens is oriented as in Fig.\ref{lens} d and f, DoF is 3.5 mm and 4.2 mm for the lens with a focal length of 20 mm and 30 mm, respectively. The simulations confirm that the selected lenses guarantee the necessary flexibility with respect to the sensor position and with acceptable losses.

Finally, the total transmission and efficiency of the optical configuration was measured at the optical exit using a continuous diode THz emitter at 270 GHz (Virginiadiode).
    \begin{figure}[h!]
        \centering
        \includegraphics[width=0.7\textwidth]{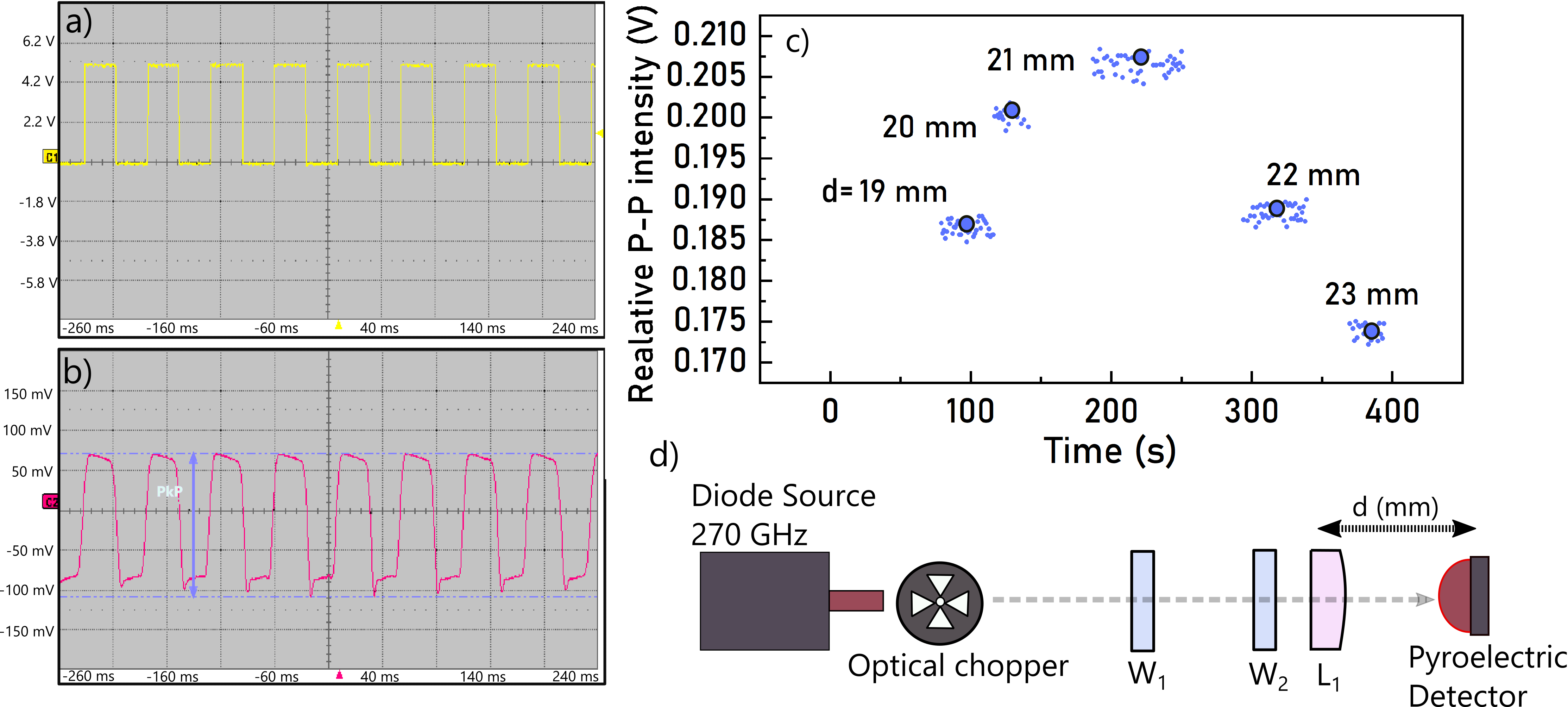}
        \caption{The overall transmission measurements on the optical path and the optimization of the lens focus on the sample positioning; a) oscilloscope screenshot of the chopper trigger signal; b) screenshot of the diode photo-response; c) the measured photo-response values from the peak to peak measurement of the amplified potential (V$_{p-p}$) of the pyroelectric sensor; d) the measurement set up with the optical line with the two windows and the 20 mm focal length lens. }
        \label{diod}
        \end{figure}
The total transmission of the optical line was measured using a calibrated pyroelectric detector with the voltage sensitivity of $\eta$=8000 V/W, positioned at the sample position in the center of the magnetic coil aperture. The incident beam was chopped at 16 Hz and sent through the optical line of the cryostat (see figure \ref{diod}d). The pyroelectric response was triggered and measured with and without the optical line, moving the final stage of the focusing lens. The lens was shifted in a 5 mm range to optimize the focal distance and the transmission. The measurement was performed looking at the peak to peak amplified voltage response of the pyroelectric detector triggered by the output frequency of the chopper (see figure \ref{diod}a,b).

The lens distance was varied via the perforated support designed for an accurate translation inside the magnetic coil aperture. The lens with the focal distance of 20 mm was used in these tests and the distance changed from d=19 to d=23 mm using the mounted screw. As can be seen in the figure \ref{diod}c, the lens shows a relatively asymmetric response while the highest intensity was measured at $f$=21 mm for the value of V$_{p-p}$=210 mV. The deviation from the maximum in the 4-5 mm distance range was measured to be of the order of the $\Delta I_d$=$\pm$ 24\%, ranging from V$_{p-p}$=175 mV to V$_{p-p}$=210mV. The result is in agreement with the simulation of the optical response of the lens. The depth of focus determined allows measurements with minor displacements of the sample positioning. The overall transmission of the optical line in the cryotube without LHe was also calculated to be 42\% in a relatively good agreement with the estimated value of $\sim$ 50\% actually, slightly lower due to the transmission difference at these wavelengths.
The cryogenic optical set up shows an acceptable response and hence could be successfully used for the investigation of new proximity array devices that require a sufficient number of photon and the control over external magnetic and electric field \cite{Rezvani2020,Rezvanion}.

\section{conclusions}
We presented here the characteristics of a compact liquid Helium immersion cryostat designed and built with an optical setup optimized for experiments using radiation at long wavelengths. The system has been specifically designed to characterize novel proximity junction array detectors and other 2D imaging detectors working at LHe temperature. The cryostat allows to maintain the sample at the temperature of 4.2 K for six hours, while it may reach the lowest temperature of 2 K upon LHe bath pumping. A liquid Nitrogen boiling chamber is used to reduce bubbling inside the cryotube.
The optical configuration of the device is made by three stages of optical elements: a Zeonex window, a crystal quartz window and a Zeonex focusing lens, out of which the latter two are immersed in liquid Helium. All materials of the optical windows were selected on the base of accurate simulations and have a thickness of about 2 mm, i.e., 2 mm for Zeonex and 2.2 mm for crystal quartz, considering both materials tensile stress and the differential pressure they are withstanding inside the cryostat as well as their optical parameters. The transmission of the Zeonex windows were measured using a two stage photo-diode configurations and for a thickness of 2 mm are characterized by a transmission of 80\% at 1 THz. Based on the optical layout, different configurations were simulated with a spatial resolution ranging from 0.48 to 1 mm and the configuration with the highest estimated resolution was realized. Furthermore, several other configurations were simulated and compared to verify the tunability of the DoF in the range 2.3 mm to 8 mm in order to compensate sample positioning misalignments.
The total transmission of the optical configuration exhibits a maximum value of 42\%  while a DoF deviation of 24\% was measured by changing the lens focal distance. These results measured in cryogenic conditions are in good agreement with theoretical estimations.
Summarizing, this original cryogenic systems matches the demanding requests of the investigation of samples and devices to be probed at low temperature, long wavelengths and when external electric and magnetic fields have to be applied.

\section{Acknowledgments}
This project has been supported within the framework of the TERA project of the V$^{th}$ Committee of the INFN.
We like to thank S. Cibella and P. Carelli for their support in the assembly of the apparatus for pyroelectric measurements and A. Grilli and A. Raco for their valuable support in the construction of the apparatus.
We acknowledge the financial support of the Bilateral Cooperation Agreement between Italy and Japan of the Italian Ministry of Foreign Affairs and of the International Cooperation (MAECI) in the framework of the project of major relevance N. PGR0072.
\section{Data Availability}
The data that supports the findings of this study are available within the article.
\bibliographystyle{unsrt}
\bibliography{SI-J-4}

\begin{thebibliography}{10}

\bibitem{Niedermayr_2014}
Michael Niedermayr, Kirill Lakhmanskiy, Muir Kumph, Stefan Partel, Jonannes
  Edlinger, Michael Brownnutt, and Rainer Blatt.
\newblock Cryogenic surface ion trap based on intrinsic silicon.
\newblock {\em New Journal of Physics}, 16(11):113068, nov 2014.

\bibitem{schwarz70}
M.~Schwarz, O.~O. Versolato, A.~Windberger, F.~R. Brunner, T.~Ballance, S.~N.
  Eberle, J.~Ullrich, P.~O. Schmidt, A.~K. Hansen, A.~D. Gingell, M.~Drewsen,
  and J.~R.~Crespo López-Urrutia.
\newblock Cryogenic linear paul trap for cold highly charged ion experiments.
\newblock {\em Review of Scientific Instruments}, 83(8):083115, 2012.

\bibitem{micke93}
P.~Micke, J.~Stark, S.~A. King, T.~Leopold, T.~Pfeifer, L.~Schmöger,
  M.~Schwarz, L.~J. Spieß, P.~O. Schmidt, and J.~R. Crespo López-Urrutia.
\newblock Closed-cycle, low-vibration 4 k cryostat for ion traps and other
  applications.
\newblock {\em Review of Scientific Instruments}, 90(6):065104, 2019.

\bibitem{andrews43}
Steven~S. Andrews and Steven~G. Boxer.
\newblock A liquid nitrogen immersion cryostat for optical measurements.
\newblock {\em Review of Scientific Instruments}, 71(9):3567--3569, 2000.

\bibitem{RN1523}
S.~S. Dhillon, M.~S. Vitiello, E.~H. Linfield, A.~G. Davies, M.~C. Hoffmann,
  J.~Booske, C.~Paoloni, M.~Gensch, P.~Weightman, G.~P. Williams,
  E.~Castro-Camus, D.~R.~S. Cumming, F.~Simoens, I.~Escorcia-Carranza,
  J.~Grant, S.~Lucyszyn, M.~Kuwata-Gonokami, K.~Konishi, M.~Koch, C.~A.
  Schmuttenmaer, T.~L. Cocker, R.~Huber, A.~G. Markelz, Z.~D. Taylor, V.~P.
  Wallace, J.~A. Zeitler, J.~Sibik, T.~M. Korter, B.~Ellison, S.~Rea,
  P.~Goldsmith, K.~B. Cooper, R.~Appleby, D.~Pardo, P.~G. Huggard, V.~Krozer,
  H.~Shams, M.~Fice, C.~Renaud, A.~Seeds, A.~Stohr, M.~Naftaly, N.~Ridler,
  R.~Clarke, J.~E. Cunningham, and M.~B. Johnston.
\newblock The 2017 terahertz science and technology roadmap.
\newblock {\em Journal of Physics D-Applied Physics}, 50(4), 2017.

\bibitem{Ortolani2006}
M.~Ortolani, S.~Lupi, L.~Baldassarre, U.~Schade, P.~Calvani, Y.~Takano,
  M.~Nagao, T.~Takenouchi, and H.~Kawarada.
\newblock {Low-Energy Electrodynamics of Superconducting Diamond}.
\newblock {\em Physical Review Letters}, 97(9):097002, aug 2006.

\bibitem{Nanni2015}
Emilio~A. Nanni, Wenqian~R. Huang, Kyung-Han Hong, Koustuban Ravi, Arya
  Fallahi, Gustavo Moriena, R.~J. {Dwayne Miller}, and Franz~X. K{\"{a}}rtner.
\newblock {Terahertz-driven linear electron acceleration}.
\newblock {\em Nature Communications}, 6(1):8486, dec 2015.

\bibitem{Jepsen2011}
P.U. Jepsen, D.G. Cooke, and M.~Koch.
\newblock {Terahertz spectroscopy and imaging - Modern techniques and
  applications}.
\newblock {\em Laser {\&} Photonics Reviews}, 5(1):124--166, jan 2011.

\bibitem{Saeedkia2013}
Daryoosh Saeedkia.
\newblock {\em {Handbook of terahertz technology for imaging, sensing and
  communications}}.
\newblock Woodhead Publishing Limited, 2013.

\bibitem{Karasik2011}
Boris~S. Karasik, Andrei~V. Sergeev, and Daniel~E. Prober.
\newblock {Nanobolometers for THz Photon Detection}.
\newblock {\em IEEE Transactions on Terahertz Science and Technology},
  1(1):97--111, sep 2011.

\bibitem{Toma2015}
Andrea Toma, Salvatore Tuccio, Mirko Prato, Francesco {De Donato}, Andrea
  Perucchi, Paola {Di Pietro}, Sergio Marras, Carlo Liberale, Remo {Proietti
  Zaccaria}, Francesco {De Angelis}, Liberato Manna, Stefano Lupi, Enzo {Di
  Fabrizio}, and Luca Razzari.
\newblock {Squeezing Terahertz Light into Nanovolumes: Nanoantenna Enhanced
  Terahertz Spectroscopy (NETS) of Semiconductor Quantum Dots}.
\newblock {\em Nano Letters}, 15(1):386--391, jan 2015.

\bibitem{DApuzzo2017}
Fausto D'Apuzzo, Alba~R. Piacenti, Flavio Giorgianni, Marta Autore,
  Mariangela~Cestelli Guidi, Augusto Marcelli, Ulrich Schade, Yoshikazu Ito,
  Mingwei Chen, and Stefano Lupi.
\newblock {Terahertz and mid-infrared plasmons in three-dimensional nanoporous
  graphene}.
\newblock {\em Nature Communications}, 8(1):14885, apr 2017.

\bibitem{Rezvani_2016_3}
S~J Rezvani, N~Pinto, E~Enrico, L~D'Ortenzi, A~Chiodoni, and L~Boarino.
\newblock Thermally activated tunneling in porous silicon nanowires with
  embedded si quantum dots.
\newblock {\em Journal of Physics D: Applied Physics}, 49(10):105104, feb 2016.

\bibitem{Rezvani2}
S.~J. Rezvani, N.~Pinto, L.~Boarino, F.~Celegato, L.~Favre, and I.~Berbezier.
\newblock Diffusion induced effects on geometry of ge nanowires.
\newblock {\em Nanoscale}, 6:7469--7473, 2014.

\bibitem{Rezvani_2016}
S~J Rezvani, N~Pinto, E~Enrico, L~D'Ortenzi, A~Chiodoni, and L~Boarino.
\newblock Thermally activated tunneling in porous silicon nanowires with
  embedded si quantum dots.
\newblock {\em Journal of Physics D: Applied Physics}, 49(10):105104, feb 2016.

\bibitem{Rezi_4}
N.~Pinto, S.~J. Rezvani, L.~Favre, I.~Berbezier, M.~Fretto, and L.~Boarino.
\newblock Geometrically induced electron-electron interaction in semiconductor
  nanowires.
\newblock {\em Applied Physics Letters}, 109(12):123101, 2016.

\bibitem{Rezvani2020}
J.~Rezvani, D.~{Di Gioacchino}, C.~Gatti, N.~Poccia, C.~Ligi, S.~Tocci,
  M.~{Cestelli Guidi}, S.~Cibella, S.~Lupi, and A.~Marcelli.
\newblock {Tunable Vortex Dynamics in Proximity Junction Arrays: A Possible
  Accurate and Sensitive 2D THz Detector}.
\newblock {\em Acta Physica Polonica A}, 137(1):17--20, jan 2020.

\bibitem{Gioacchino2017}
Daniele~Di Gioacchino, Nicola Poccia, Martijn Lankhorst, Claudio Gatti, Bruno
  Buonomo, Luca Foggetta, Augusto Marcelli, and Hans Hilgenkamp.
\newblock {A Novel Particle/Photon Detector Based on a Superconducting
  Proximity Array of Nanodots}.
\newblock {\em Journal of Superconductivity and Novel Magnetism},
  30(2):359--363, feb 2017.

\bibitem{Poccia2015}
Nicola Poccia, Tatyana~I. Baturina, Francesco Coneri, Cor~G. Molenaar,
  X.~Renshaw Wang, Ginestra Bianconi, Alexander Brinkman, Hans Hilgenkamp,
  Alexander~A. Golubov, and Valerii~M. Vinokur.
\newblock {Critical behavior at a dynamic vortex insulator-to-metal
  transition}.
\newblock {\em Science}, 349(6253):1202--1205, sep 2015.

\bibitem{Tera}
Tera project, lnf, infn.
\newblock \url{ http://dr.lnf.infn.it/technological-research/}.

\bibitem{Castro-Camus16}
E.~Castro-Camus and M.~Alfaro.
\newblock Photoconductive devices for terahertz pulsed spectroscopy: a review.
\newblock {\em Photon. Res.}, 4(3):A36--A42, Jun 2016.

\bibitem{Ferguson2002}
Bradley Ferguson and Xi-Cheng Zhang.
\newblock {Materials for terahertz science and technology}.
\newblock {\em Nature Materials}, 1(1):26--33, sep 2002.

\bibitem{Zhang2010}
Xi-Cheng Zhang and Jingzhou Xu.
\newblock {\em {Introduction to THz Wave Photonics}}.
\newblock Springer US, Boston, MA, 2010.

\bibitem{anna}
A.~D’Arco, M.~Di~Fabrizio, M.~Dolci, V.and~Petrarca, and S.~Lupi.
\newblock Thz pulsed imaging in biomedical applications.
\newblock {\em Condensed matter}, 5:25, 2020.

\bibitem{Cha2006}
Hyuk~Jin Cha, Young~Uk Jeong, Seong~Hee Park, Byung~Cheol Lee, and Seung~Han
  Park.
\newblock {Transmission-property measurements of crystal quartz and plastic
  material in the terahertz region}.
\newblock {\em Journal of the Korean Physical Society}, 49:354--358, 2006.

\bibitem{Born2000}
Max Born, Emil Wolf, and Eugene Hecht.
\newblock {Principles of Optics: Electromagnetic Theory of Propagation,
  Interference and Diffraction of Light}.
\newblock {\em Physics Today}, 53(10):77--78, oct 2000.

\bibitem{tydex}
Tydex optics.
\newblock \url{http://www.tydexoptics.com}.

\bibitem{comsol}
Comsol multiphysics.
\newblock \url{http://www.comsol.com}.

\bibitem{Tofani2019}
Silvia Tofani, Dimitrios~C. Zografopoulos, Mauro Missori, Renato Fastampa, and
  Romeo Beccherelli.
\newblock {Terahertz focusing properties of polymeric zone plates characterized
  by a modified knife-edge technique}.
\newblock {\em Journal of the Optical Society of America B}, 36(5):D88, may
  2019.

\bibitem{Tofani20192}
Silvia Tofani, Dimitrios~C. Zografopoulos, Mauro Missori, Renato Fastampa, and
  Romeo Beccherelli.
\newblock {High-Resolution Binary Zone Plate in Double-Sided Configuration for
  Terahertz Radiation Focusing}.
\newblock {\em IEEE Photonics Technology Letters}, 31(2):117--120, jan 2019.

\bibitem{Rezvanion}
S.J. Rezvan, D.~Di Gioacchino, C.~Gatti, C.~Ligi, M.~Cestelli Guidi,
  S.~Cibella, M.~Fretto, N.~Poccia, S.~Lupi, and A.~Marcelli.
\newblock Proximity array device: a novel photon detector working in long
  wavelengths.
\newblock {\em Condensed matter}, 2020.

\end{thebibliography}
\end{document}